\documentstyle[12pt,aaspp4,psfig]{article}

\lefthead{left}
\righthead{right}

\newcommand{\gtilde}
 {~ \raisebox{-1ex}{$\stackrel{\textstyle >}{\sim}$} ~}
\newcommand{\ltilde}
 {~ \raisebox{-1ex}{$\stackrel{\textstyle <}{\sim}$} ~}

\begin{document}

%\vspace*{-1cm}
%\hspace*{12cm}
%RESCEU-14/97 \\
%\hspace*{12cm}

\title{Evolution of the Luminosity Density in the Universe: Implications
for the Nonzero Cosmological Constant}

\author{T. Totani \altaffilmark{1}, 
        Y. Yoshii \altaffilmark{2, 3}, 
       \& K. Sato \altaffilmark{1, 3}} 

\altaffiltext{1}{Department of Physics, School of Science, 
the University of Tokyo, Tokyo 113, Japan \\
E-mail: totani@utaphp2.phys.s.u-tokyo.ac.jp}

\altaffiltext{2}{Institute of Astronomy, Faculty of Science, 
the University of Tokyo, 2-21-1 Osawa, Mitaka, Tokyo 181, Japan}

\altaffiltext{3}{Research Center for the Early Universe, School of Science, 
the University of Tokyo, Tokyo 113, Japan 
}

%\begin{center}
%\it To Apper in ApJ Letters
%\end{center}

\begin{abstract}
We show that evolution of the luminosity density of galaxies in the 
universe provides a powerful test for the geometry of the universe.  
Using reasonable galaxy evolution models of population synthesis which 
reproduce the colors of local galaxies of various morphological types, 
we have calculated the luminosity density of galaxies as a function of 
redshift $z$.  Comparison of the result with recent measurements by the 
Canada-France Redshift Survey in three wavebands of 2800{\AA}, 4400{\AA}, 
and 1$\mu$m at $z<1$ indicates that the $\Lambda$-dominated flat universe 
with $\lambda_0\sim0.8$ is favored, and the lower limit on $\lambda_0$ 
yields 0.37 (99\% C.L.) or 0.53 (95\% C.L.) if $\Omega_0+\lambda_0=1$.  
The Einstein-de Sitter universe with $(\Omega_0, \lambda_0)=(1, 0)$ and 
the low-density open universe with (0.2, 0) are however ruled out with 
99.86\% C.L. and 98.6\% C.L., respectively.  The confidence levels quoted
apply unless the standard assumptions on galaxy evolution are drastically 
violated. 

We have also calculated a global star formation rate in the universe to 
be compared with the observed rate beyond $z \sim 2$.  We find from this 
comparison that spiral galaxies are formed from material accretion over an 
extended period of a few Gyrs, while elliptical galaxies are formed from 
initial star burst at $z\gtilde 5$ supplying enough amount of metals and 
ionizing photons in the intergalactic medium.

\end{abstract}


\section{Introduction}
Recent progress in deep redshift surveys of galaxies reaching $z \sim 1$ 
gives valuable information on the history of galaxy evolution 
(Lilly et al. 1995; Ellis et al. 1996).  The Canada-France Redshift Survey 
(CFRS) consisting of 730 $I_{AB}$-selected galaxies with $17.5<I_{AB}<22.5$ 
revealed a marked evolution in their comoving luminosity density in three 
wavebands of 2800{\AA}, 4400{\AA}, and 1$\mu$m in the range of $z$ = 0--1 
(Lilly et al. 1996). They argued that their data can be explained by a galaxy
evolution model of population synthesis with star formation rate proportional 
to $\tau ^{-2.5}$ where $\tau$ is the elapsed time from the initial turn-on 
of star formation.  In practice, however, their use of a single index 
$-2.5$ for all galaxies is  not justified because the present-day colors 
of galaxies are strongly correlated with their morphological type through 
type-dependent history of star formation (Tinsley 1980), and a single-type 
analysis is known to produce a bias in interpreting the data when applied 
to a sample with a realistic mixture of morphological types 
(Yoshii \& Takahara 1989).  One therefore has to construct type-dependent 
evolution models of galaxies and sum up with relative proportion of their 
types.  If the global history of star formation for the composite of all 
types of galaxies is determined in this way, cosmological models can then 
be tested against the observations of the luminosity density evolution in 
the universe. 

In this {\it Letter}, we calculate an evolution of the luminosity density 
using 
the galaxy evolution models of population synthesis developed by Arimoto 
\& Yoshii (1987, hereafter AY) and Arimoto, Yoshii, \& Takahara (1992, 
hereafter AYT) where the effect of chemical evolution is appropriately 
taken into account.  Since the integrated color of stars in a galaxy is 
very sensitive to an assumed history of star formation, the present-day 
spectral energy distribution (SED) in the ultraviolet through near-infrared 
region for each type of galaxies can only be reproduced by almost identical 
luminosity evolution as a function of look-back time until the epoch 
corresponding to an average age of stars in a galaxy (AY, AYT).  
These models are therefore reliable especially at low redshifts
of $z<1$ (Yoshii \& Takahara 1988) where we constrain the cosmological
parameters from the CFRS data.  A global rate of star formation is also 
calculated and is used to discuss an early evolution of galaxies from some 
recent observations of star formation rate at $z>1$.

\section{Theoretical Model of the Luminosity Density Evolution}
Let ${\cal L}(t, \lambda)$ be the comoving luminosity density of galaxies 
measured at time $t$ from the epoch of galaxy formation and at rest-frame 
wavelength $\lambda$ per unit wavelength.  Then ${\cal L}(t, \lambda)$ is 
expressed as, after summing over galaxy types, 
\begin{equation}
{\cal L}(t, \lambda) = \sum_k \int dL_B \psi_k(L_B) L_k(t,\lambda;L_B) \ ,
\end{equation}
where $\psi_k(L_B)$ is the present-day luminosity function for blue-selected
galaxies of $k$-th type and $L_k(t,\lambda;L_B)$ is the luminosity at $t$ 
and $\lambda$ per unit wavelength for a galaxy of $k$-th type whose 
present-day $B$-band luminosity is $L_B$.   Assuming that the SED evolution 
$f_k(t,\lambda)$ depends only on the galaxy type and not on the luminosity 
of each galaxy, we write 
\begin{equation}
L_k(t,\lambda;L_B) = \left[ \frac{f_k(t, \lambda)}
{f_k(T_G,\lambda_B)} \right] L_k(T_G,\lambda_B;L_B) \ ,
\end{equation}
where $T_G$ is the present age of galaxies and $\lambda_B$=4400{\AA}. 
We note that the assumption of luminosity-independent evolution is 
consistent with the method of evaluating the luminosity density from the 
CFRS data where only galaxies brighter than $\sim L^*$ are sampled at 
$z\gtilde 0.5$ and the luminosity function is extrapolated 
to fainter magnitudes with its shape kept constant (Lilly et al. 1996).  
Thereby, the luminosity density at $t$ and $\lambda$ can be related to the 
present-day $B$-band luminosity density of each type:
\begin{equation}
{\cal L}(t,\lambda) = \sum_k \left[ \frac{f_k(t,\lambda)}
{f_k(T_G,\lambda_B)} \right] {\cal L}_k(T_G,\lambda_B) \ .
\end{equation}  
We classify galaxies into E/S0, Sab, Sbc, Scd, and Sdm which evolve with
different star formation histories.  The evolving SEDs of spiral galaxies 
are given in AYT over a range of wavelength spanning from  1550 to 
34000{\AA}, and those for elliptical galaxies are taken from the 
$10^{12}M_\odot$ (baryon mass) model in AY for 3600--34000{\AA}.  
The ultraviolet SED of elliptical galaxies, the origin of which is still 
a matter of debate, is prescribed in the same way as in Yoshii \& Peterson 
(1991), but the following results are hardly changed by this uncertain 
UV-prescription because E/S0 galaxies contribute little to the total 
luminosity density at $\lambda<3600${\AA} as long as redshifts of $z<1$ are 
concerned in our analysis.

In the galaxy evolution models,
stars are assumed to be formed at a rate proportional to a power of
mass fraction of gas in a galaxy such as $C(t)=\nu f^n(t)$.
The rate coefficient $\nu$ is chosen in a way to 
reproduce the present-day colors of each galaxy type assuming $T_G$=15 Gyrs 
and the power index $n$ is generally believed to lie in the range of 1--2 
(see, e.g., Kennicutt 1989).   We consider four different evolution models 
for spiral galaxies such as S1, S2, I1, and I2 (for details see AYT).  
The symbol `S' refers to the simple, closed-box model, while `I' to the 
infall model which allows for material infall into the disk region.  
The number attached to `S' and `I' is the adopted value of power index 
$n$ in $C(t)$.  The above four models are considered to cover a range of 
possible star formation histories and are used to assess the uncertainty 
associated with the prescriptions of galaxy evolution in our analysis.

\begin{figure}
  \begin{center}
    \leavevmode\psfig{figure=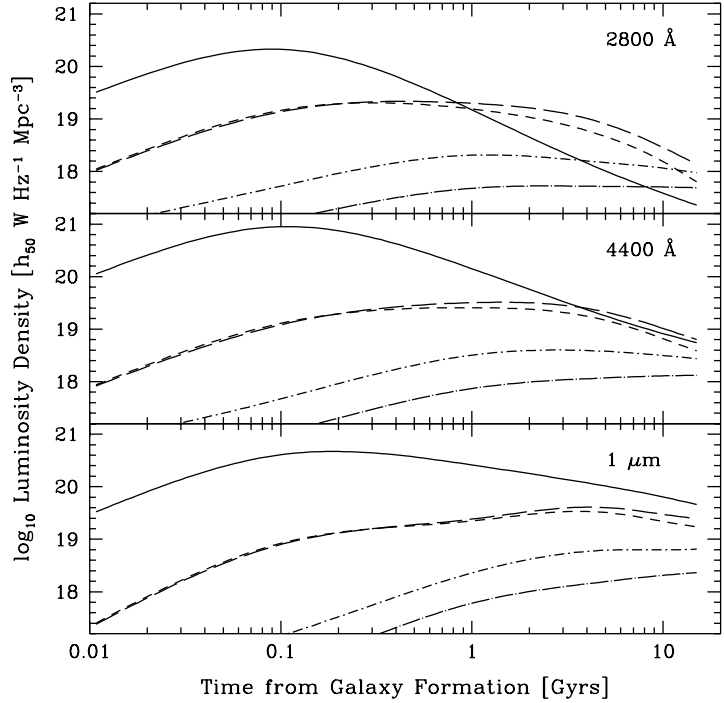,width=10cm}
  \end{center}
\caption{Evolution of the comoving luminosity density at 2800{\AA}, 
4400{\AA} and 1$\mu$m as a function of time for model galaxies of different 
morphological types.  The galactic wind model from AY is used for E/S0 
(solid line), and the S1 models from AYT for spiral galaxies of Sab 
(short-dashed), Sbc (long-dashed), Scd (dot-short-dashed), and Sdm 
(dot-long-dashed).  The luminosity density at 15 Gyrs for the composite
of all types is normalized to
${\cal L}(4400{\rm \AA})=10^{19.296} h_{50}$ W Hz$^{-1}$ Mpc$^{-3}$,
and the relative contribution among the types is adopted from the CfA mix 
in Table 1. 
}
\label{fig:l-dens-t}
\end{figure}

In order to compare with the CFRS data, we set the normalization as 
${\cal L}(T_G,\lambda_B)=10^{19.296}h_{50}$ W Hz$^{-1}$ Mpc$^{-3}$, where 
$h_{50}\equiv H_0/$(50 km/s/Mpc), following Lilly et al. (1996). 
The contribution from each morphological type in ${\cal L}(T_G,\lambda_B)$ is 
evaluated from the Schechter parameters $(\phi^*_k,L_{B,k}^*,\alpha_k)$
determined by Efstathiou, Ellis, \& Peterson (1988) for the Anglo-Australian
Redshift Survey (AARS), the 
Kirshner-Oemler-Schechter survey (KOS),
the Revised Shapley-Ames sample (RSA), and 
the Center for Astrophysics redshift survey (CfA).
The Schechter parameters, converted into the same $B$-band system for the 
four redshift surveys, are summarized in Table 1 of Yoshii \& Takahara (1989)
and are used to calculate the proportion of local luminosity density among 
galaxy types as ${\cal L}_k=\phi^*_k L_k^* \Gamma(\alpha_k + 2)$.
The results are tabulated in Table \ref{table:l-dens-ratio}. 
When spiral galaxies are not classified into the four subclasses in the AARS,
KOS, and RSA surveys, the relative proportion among the subclasses is
assumed to be the same as that for the CfA survey.  
Fig. \ref{fig:l-dens-t} shows the luminosity density
evolution as a function of time for each galaxy type assuming the CfA
mix and $T_G$=15 Gyrs. In this figure the S1 models are used for
the evolution of spiral galaxies.  
Unless otherwise stated, the CfA mix is used below 
as a standard choice.  

\section{Probing the Cosmological Parameters}
We consider three representative cosmological models: the Einstein-de Sitter
(EdS) universe with 
($\Omega_0,\lambda_0)=(1,0)$, an open universe with (0.2,0), and 
a $\Lambda$-dominated, flat universe with (0.2, 0.8).  The redshift of
galaxy formation is assumed to be $z_F = 5$.  Different values of $z_F$
hardly change our result as far as we restrict ourselves in $z<1$
(see discussion in \S \ref{section:discussion}).  
The Hubble constant $h\equiv H_0/$(100 km/s/Mpc) is taken as 0.5, 0.6, 
and 0.7 for the three cosmological models in order to give a
reasonable age of galaxies (12.1, 12.4, and 13.6 Gyrs, respectively).
Although the galaxy evolution models reproduce the present-day SED at 
$T_G$=15 Gyrs, we can use these models for the case of $T_G$=10--15 Gyrs 
as well, because the luminosity evolution becomes very weak and 
changes the SED very little after $t\sim$10 Gyrs. 

\begin{figure}
  \begin{center}
    \leavevmode\psfig{figure=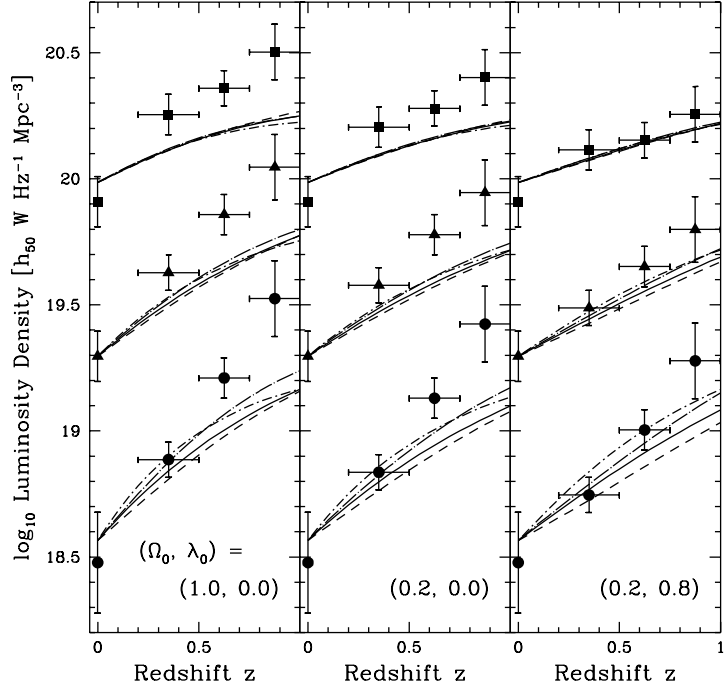,width=10cm}
  \end{center}
\caption{Evolution of the comoving luminosity density as a function of 
redshift for the composite of all galaxy types, shown for the 
Einstein-de Sitter 
universe with $(\Omega_0,\lambda_0)=(1,0)$ ({\it left panel}), 
the open universe with (0.2, 0) ({\it middle panel}), and the 
$\Lambda$-dominated flat universe with (0.2, 0.8) ({\it right panel}).
The CFRS data points (Lilly et al. 1996), 
scaled to $h=0.5$, are shown by circles (2800{\AA}), 
triangles (4400{\AA}), and squares (1$\mu$m).  The four theoretical curves 
for each of 2800{\AA}, 4400{\AA}, and 1$\mu$m correspond to the evolution 
models of S1 (solid line), S2 (dashed), I1 (dot-short-dashed), and I2 
(dot-long-dashed) for spiral galaxies.  The galactic wind model for 
elliptical galaxies is used in common.  The curves are normalized to
coincide with the 4400 {\AA} data at $z$ = 0.
}
\label{fig:l-dens-z}
\end{figure}

In order to compare with the CFRS data obtained for $z<1$, we transform
the theoretical predictions into those expressed in terms of $z$.
In Fig. \ref{fig:l-dens-z}, we show the comoving luminosity densities
at 2800{\AA}, 4400{\AA}, and 1$\mu$m as a function of $z$ for the EdS
universe ({\it left panel}), the open universe ({\it middle panel}), and
the $\Lambda$-dominated universe ({\it right panel}).  The CFRS data 
points are those from the ``luminosity-function-estimated'' values
(Lilly et al. 1996)
for 2800{\AA} (circles), 4400{\AA} (triangles), and 1 $\mu$m (squares). 
The four theoretical curves for each waveband correspond to the 
variation of the evolution models of spiral galaxies:
S1(solid), S2(dashed), I1(dot-short-dashed), and 
I2(dot-long-dashed).  In this figure all the predicted and observed
luminosity densities are shown after scaled to $h = 0.5$, and the
theoretical curves are made to agree with the 4400 {\AA} data at $z=0$.

The behaviors of theoretical curves are not very different among the 
three cosmological models because the elapsed time between $z=0$ and 1 is 
roughly the same as 8 Gyrs.  On the other hand, the observed luminosity 
densities, which have been derived from the direct observables of apparent 
luminosities and redshifts of galaxies, generally depend on the 
cosmological parameters.   It is clear from Fig. \ref{fig:l-dens-z} that
the observed ${\cal L}$-evolution at $z=0-1$ becomes flatter with decreasing 
$\Omega_0$ or increasing $\lambda_0$, and eventually falls in agreement 
with the $\Lambda$-dominated universe.  In sharp contrast, however, the 
observed ${\cal L}$-evolution is too steep to agree with the EdS or open 
universe, and such a discrepancy is much more considerable at 1$\mu$m.
We note that the ${\cal L}$-evolution is reliably modeled  at 1$\mu$m,
offering an ideal cosmological test when compared to shorter wavelengths.
This is because the luminosity density at 1$\mu$m is always dominated by 
ellipticals and is quite insensitive to details of recent star formation 
history in spiral galaxies, as shown in the lower panel of Fig.1.

For the purpose of quantitative comparison, we have calculated the slope 
index of the ${\cal L}$-evolution, $\alpha \equiv d\log{\cal L}/d\log(1+z)$, 
averaged over the range of $z=0-1$ for each of the galaxy evolution models
(S1--I2).   We have repeated calculations with type mixes other than the 
CfA mix in Table \ref{table:l-dens-ratio}, but the resulting $\alpha$ is 
changed at most by 0.08.   Lilly et al. (1996) estimated the observed 
slope for the EdS cosmology as $\alpha = 3.90 \pm 0.75 (2800{\rm \AA})$, 
$2.72 \pm 0.5 (4400{\rm \AA})$, and $2.11 \pm 0.5$ (1 $\mu$m).   Using these 
observational errors which do not depend on $\Omega_0$ and $\lambda_0$, 
we have performed a $\chi^2$ analysis for $\alpha$ and found that the EdS 
and open cosmologies are inconsistent with the data with 99.86\% C.L. and 
98.6\% C.L., respectively.  Assuming a flat universe ($\Omega_0+\lambda_0=1$)
with $h=0.6$, the lower limit on $\lambda_0$ is obtained as 0.37 (99\% C.L.) 
or 0.53 (95\% C.L.).  Here, one of the galaxy evolution models (S1--I2) is 
used that gives the most conservative result in the $\chi^2$ analysis.    

\section{Discussion}
\label{section:discussion}
Our conclusion supporting the $\Lambda$-dominated universe is drawn from
using the AY and AYT models of galaxy evolution based on fairly standard 
assumptions on galaxy evolution and local properties of galaxies.   However, 
some other scenarios of galaxy evolution have been proposed so far mainly 
to solve the so-called ``excess'' faint blue galaxies around $B\sim24$.  
These include a scenario of widespread merging of galaxies (Broadhurst, 
Ellis, \& Glazebrook 1992) and a scenario which invokes an exotic 
intermediate-redshift population of dwarf galaxies which have ever faded 
away (Babul \& Rees 1992).  It is worth emphasizing that our analysis in 
this {\it Letter} is based on the luminosity density of galaxies but not 
on their number density.  
Luminosity density is generally less sensitive to mergers than number 
density, because increase of number of galaxies at high redshifts
is compensated by the corresponding decrease of luminosity of individual
galaxies.  The exotic dwarf galaxies may significantly contribute in the 
number density, but it is doubtful that such a faint population occupies a 
large fraction in the CFRS galaxies at $0.5\ltilde z\ltilde 1$.  In fact, 
their HST images exhibit the same range of morphological types as seen 
locally and the CFRS galaxies at $z\sim$0.75 are broadly similar to the 
local galaxy population (Schade et al. 1995), indicating that exotic 
evolutionary processes do not dominate in luminous galaxies. 

\begin{figure}
  \begin{center}
    \leavevmode\psfig{figure=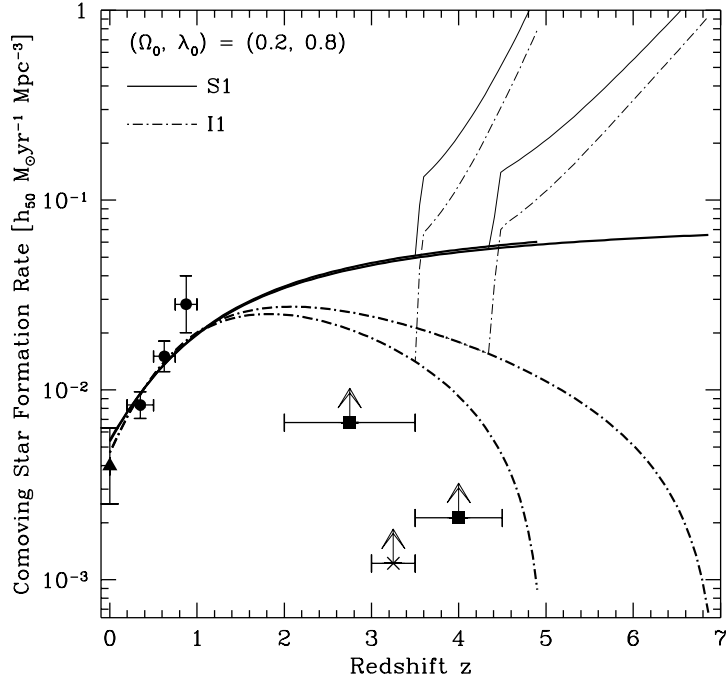,width=10cm}
  \end{center}
\caption{Star formation rate (SFR) in the universe as a function of 
redshift.  The $\Lambda$-dominated flat universe with 
$(\Omega_0,\lambda_0)=(0.2,0.8)$ is assumed for both predicted and
observed SFRs shown in this figure after scaled to $h=0.5$.   The data 
points are those compiled by Madau et al. (1996). 
The theoretical curves are based on the S1 (solid)
or I1 (dot-dashed) models for spiral galaxies 
and the galactic wind model for elliptical galaxies. 
Galaxies of all types are assumed to be formed at $z_F=5$ or 7, with 
$h = 0.7$. The SFRs from spiral galaxies only
are shown by thick lines, while those from all types including elliptical
galaxies by thin lines.   The theoretical curves are normalized to agree
with the data at $z<1$.
}
\label{fig:sfr-history}
\end{figure}

The redshift of galaxy formation, though still uncertain, has been assumed 
to be $z_F=5$, because different choices give a convergent result below 
$z\sim 1$ and do not affect the cosmological test presented in this 
{\it Letter}.  Recent progress in direct observations of star-forming 
galaxies at $z\gtilde 2$ however provides more useful constraint on $z_F$.  
In Fig. \ref{fig:sfr-history}, assuming the $\Lambda$-dominated universe 
with $(\Omega_0,\lambda_0)=(0.2,0.8)$, we plot the observed star formation
rates (SFRs) in the universe which are compiled by Madau et al. (1996),
together with the predicted SFRs with $z_F=5$ and 7 assuming $h = 0.7$ 
(for details see Totani, Sato, \& Yoshii 1996).   The SFRs, observed or
predicted, are shown after scaled to $h=0.5$ in this figure.  Thick lines 
show the SFRs from spiral galaxies, and thin lines from all types including 
elliptical galaxies.  Since the absolute SFRs are uncertain by a factor of 
2--3, we normalize the curves to agree with the data at $z<1$.  Inspection 
of this figure clearly indicates that $z_F$ is larger than at least $\sim 4$.
According to Madau et al. (1996), the observed SFRs beyond $z$ = 1 shown in 
Fig. \ref{fig:sfr-history} should be considered as lower limits to the real 
values.  However, since their analysis is based on a sample which covers a 
significant range of magnitude in the galaxy luminosity function, it may be 
safe to assume that these lower limits are not very different from the real 
values.  If this is the case, the infall models of spiral galaxies with 
$z_F\sim$ 4--5 well explain the SFR trend at $z\sim$ 2--4, whereas the 
model of elliptical galaxies overpredicts the SFRs unless they were formed 
at $z_F\gtilde 5$.

We point out that elliptical galaxies having high SFRs release metals 
out of the system into the intergalactic medium (IGM) through galactic winds.
Miralda-Escude and Rees (1997) argued that the average metal abundance of 
$\bar Z\sim 2\times 10^{-4}$, which is observed in the Ly$\alpha$ forests 
seen in quasar spectra, needs one Type II supernova per 5000$M_\odot$ 
baryons in the IGM and these supernova progenitors should 
have emitted about 20 ionizing photons per each baryon, which are enough to 
reionize the universe.   In our models, an elliptical galaxy with 
$M_{\rm baryon}=10^{12}M_\odot$ produces about 2.5 $\times 10^{10}$ 
Type II supernovae during the first 0.7 Gyr (Totani, Sato, \& Yoshii 1996) 
and then releases metal-enriched gas of about 7 \% of $M_{\rm baryon}$ into the
IGM (AY).  
If the baryon density in the universe is taken as a typical value of 
$\Omega_b h^2=0.02$ from the Big-Bang nucleosynthesis, the average metal 
abundance in the universe is given by $\bar Z\sim 1.7\times 10^{-4}h^2$.
Since this value agrees with the estimate by Miralda-Escude and Rees (1997), 
we can regard the elliptical galaxies as a source of supplying not only 
the metals in the IGM but also the ionizing photons which reionize the 
universe.

\acknowledgements
We would like to thank an anonymous referee for useful comments.
This work has been supported in part by the Grant-in-Aid for the 
Center-of-Excellence (COE) Research (07CE2002) and for the Scientific 
Research Fund (05243103, 07640386, 3730) of the Ministry of Education, 
Science, and Culture in Japan.

%%%%%%%%%%%%%%%%%%%%%%%%%%%%%%%% TABLES %%%%%%%%%%%%%%%%%%%%%%%%%%%%%%
\begin{table}
  \caption{Local Luminosity Density Proportion in Galaxy Types (4400{\AA})}
  \label{table:l-dens-ratio}
  \begin{center}
    \begin{tabular}{lccccc} \hline \hline
      & \multicolumn{5}{c}{Relative proportion} \\
      \cline{2-6}
      Sample & E/S0  & Sab & Sbc & Scd & Sdm \\
      \hline
      AARS...& 0.14 & \multicolumn{4}{c}{0.86} \\
      KOS.....& 0.21 & \multicolumn{4}{c}{0.79} \\
      RSA.....& 0.20 & \multicolumn{2}{c}{0.54} &
                       \multicolumn{2}{c}{0.26} \\
      CfA......& 0.28 & 0.19 & 0.32 & 0.14 & 0.066 \\
      \hline 
    \end{tabular}
  \end{center}
\end{table}

%%%%%%%%%%%%%%%%%%%%%%%% REFERENCES %%%%%%%%%%%%%%%%%%%%%%%%%

%%%%%%%%%%%%%%%%%%%%%% FIGURES %%%%%%%%%%%%%%%%%%%%%%%%%%%%

\newpage

\end{document}